\documentclass[aps,prl,superscriptaddress,preprintnumbers,twocolumn,amsmath,amssymb,floatfix]{revtex4-1}
\usepackage{graphicx}
\usepackage{color}
\definecolor{RED}{rgb}{1,0,0}
\definecolor{BLUE}{rgb}{0,0,1}
\begin{document}

\title{Orbital-selective Kondo entanglement and antiferromagnetic order in USb$_2$}
\author{Q. Y. Chen}
\author{X. B. Luo}
\author{D. H. Xie}
\affiliation{Science and Technology on Surface Physics and Chemistry Laboratory, Mianyang 621908, China}
\author{M. L. Li}
\affiliation{Institute of Applied Physics and Computational Mathematics, Beijing 100088, China}
\author{X. Y. Ji}
\author{R. Zhou}
\affiliation{Science and Technology on Surface Physics and Chemistry Laboratory, Mianyang 621908, China}
\author{Y. B. Huang}
\affiliation{Shanghai Institute of Applied Physics, CAS, Shanghai, 201204, China}
\author{W. Zhang}
\author{W. Feng}
\author{Y. Zhang}
\author{X. G. Zhu}
\author{Q. Q. Hao}
\author{Q. Liu}
\author{L. Huang}
\affiliation{Science and Technology on Surface Physics and Chemistry Laboratory, Mianyang 621908, China}
\author{P. Zhang}
\affiliation{Institute of Applied Physics and Computational Mathematics, Beijing 100088, China}
\author{X. C. Lai}
\affiliation{Science and Technology on Surface Physics and Chemistry Laboratory, Mianyang 621908, China}
\author{Q. Si}
\email{qmsi@rice.edu}
\affiliation{Department of Physics and Astronomy, Rice University, Houston, TX 77005, USA}
\author{S. Y. Tan}
\email{tanshiyong@caep.cn}
\affiliation{Science and Technology on Surface Physics and Chemistry Laboratory, Mianyang 621908, China}

\begin{abstract}
In heavy-fermion compounds, the dual character of $f$ electrons  underlies their rich and often exotic properties like fragile heavy quasipartilces, variety of magnetic orders and unconventional superconductivity. 5$f$-electron actinide materials  provide a rich setting to elucidate the larger and outstanding issue of the competition between magnetic order and Kondo entanglement and, more generally, the interplay among different channels of interactions in correlated electron systems.
Here, by using angle-resolved photoemission spectroscopy, we present detailed electronic structure of USb$_2$ and observed two different kinds of nearly flat bands in the antiferromagnetic state of USb$_2$. Polarization-dependent measurements show that these electronic states are derived from 5$f$ orbitals with different characters; in addition,  further temperature-dependent measurements reveal that one of them is driven by the Kondo correlations between the 5$f$ electrons and conduction electrons, while the other reflects the dominant role of the magnetic order. Our results on the low-energy electronic excitations of USb$_2$ implicate orbital selectivity as an important new ingredient for the competition between Kondo correlations and magnetic order and, by extension, in the rich landscape of quantum phases for strongly correlated $f$ electron systems.



\end{abstract}

\maketitle

Heavy fermion systems represent a prototype setting to study the physics of strong correlations, including the variety of quantum phases and their transitions \cite{Coleman.05,Si.10}. While most attention has been directed to the $4f$-electron-based rare-earth compounds,  the actinide-based compounds also have rich physics but are less studied. Uranium-based materials take a special place in this category. They  display intriguing and attractive properties, such as heavy-fermion states, unconventional superconductivity, hidden order and multiple orders \cite{Pfleiderer.09,Ott.83,Arko.84,Palstra.85,He.92,Saxena.00,Aoki.04,Fujimori.07}. It is generally believed that these properties mainly originate from the interplay between partially filled shell of 5$f$ orbitals and  broad bands of conduction electrons. Importantly, 5$f$ electrons have an intermediate character between localized 4$f$ electrons of rare-earth compounds and itinerant 3$d$ electrons of transition metals, and this dual character may be responsible for the wide variety of physical properties of uranium-based materials  \cite{Pfleiderer.09}. The $f$ electrons behave as atomic local moments at high temperature, which interact with the conduction electrons ($c$-$f$ hybridization) to produce low-energy $f$-based electronic  excitations as the temperature is decreased \cite{Shim.07,Zhu.13}. To elucidate this process, it is important to understand how the dual character of the 5$f$ electrons manifests itself in the low-energy electronic structure and physical properties of the uranium-based materials.

Another outstanding issue in the uranium-based compounds is the origin of the magnetic order, especially given that the 5$f$ electrons show dual character. A useful way to address this issue is to study the electronic structure across the magnetic transition by angle-resolved photoemission spectroscopy (ARPES). However, up to now this has only been performed in  the ferromagnetic uranium-based compound UTe\cite{Durakiewicz.04} and antiferromagnetic compound UN \cite{Fujimori.12}. Moreover, a long-standing important issue is whether $f$-electrons are itinerant or localized when the magnetic order occurs.  In a class of Ce-based 4$f$-electron materials, there has been considerable evidence that Kondo entanglement is destroyed in the antiferromagnetic order \cite{Paschen.04,Shishido.05}. It has also been emphasized that such a Kondo destruction still allows for the onset of hybridization gap \cite{Kirchner.19}, as has recently been found in the magnetically ordered phase of CeRhIn$_5$ \cite{Haze.18}. It is timely to address the interplay between Kondo entangelment and magnetic order in 5$f$-electron systems.

Antiferromagnetic USb$_2$ provides an ideal platform for such studies, as it is a moderately correlated electron system with a quasi-2D electronic structure \cite{Leciejewicz.67,Aoki.00,Kato.04,Baek.10}.
USb$_2$ crystallizes in the tetragonal structure of anti-Cu$_2$Sb type, which orders antiferromagnetically below a high  N$\acute{иж}$eel temperature of 203 K \cite{Aoki.99,Wawryk.06}. Magnetic moments of U ions are aligned ferromagnetically in the (001) planes, which are stacked along the [001] direction in an antiferromagnetic sequence ($\uparrow$$\downarrow$$\downarrow$$\uparrow$).
Previous pioneering ARPES studies on USb$_2$ have reported its bulk electronic structure by soft x-ray ARPES \cite{Fujimori.16}, and observed a narrow heavy $f$-electron band below the Fermi level ($E_F$) and the first kink structure of actinide materials  \cite{Arko.97,Takahashi.02,Baek.10,Guziewicz.04,Durakiewicz.08,Durakiewicz.14}. However, detailed electronic structure evolution across the antiferromagnetic transition is still lacking.
\begin{figure}[tbp]
\includegraphics[width=87mm]{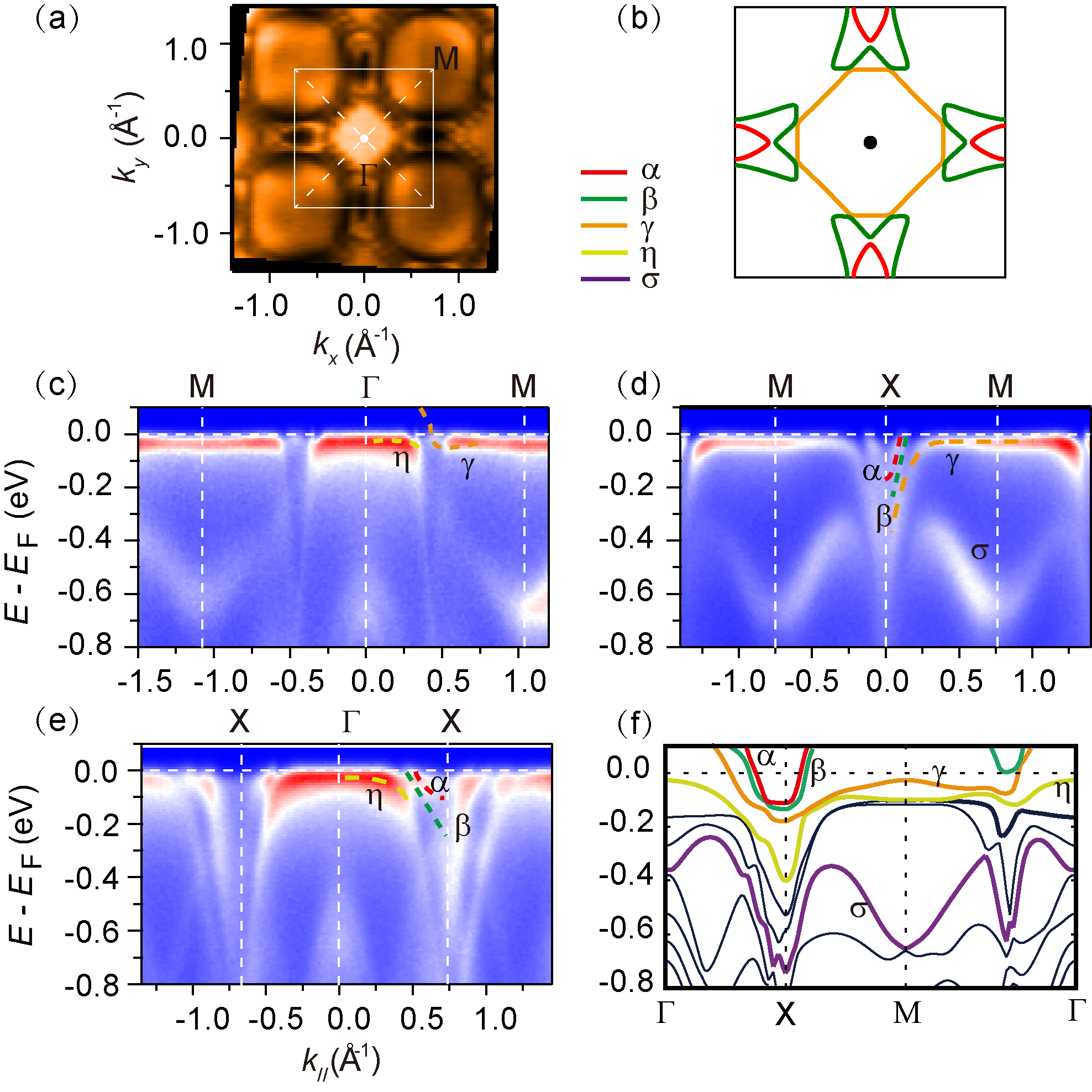}
\caption{Fermi surface and band structure of USb$_2$ taken with 102 eV photos at 20 K with LH- polarized light. (a) Photoemission intensity map of USb$_2$ at $E_F$. The intensity is integrated over a window of [$E_F$ -20 meV, $E_F$+20 meV]. Fermi surface contours are drawn with respective colors. (b) Calculated Fermi surface of the Sb-terminated surface. (c-d) Photoemission intensity distributions along (c) $\Gamma$-M, (d)M-X, and (e) $\Gamma$-X.  (f) Calculated band structure of the Sb-terminated surface.
}
\label{FS}
\end{figure}

In the present study, we report the Fermi surface topology and band structure of USb$_2$ by ARPES. Two nearly flat bands can be observed around the $\Gamma$  and M  points at low temperature, which are mainly from the 5$f$ electrons. Polarization-dependent measurements reveal that these two nearly flat bands have different orbital character. Temperature-dependent measurements have further shown that the nearly flat band around the $\Gamma$  point is due to the hybridization between the 5$f$ electrons and conduction electrons. A gap opens below the antiferromagnetic transition temperature (203 K) around the M  point, which indicates that the other nearly flat band  $\gamma$ around the M  point originates from the magnetic order. Our results reveal that  Kondo interaction can coexist with the magnetic order by manifesting themselves in different bands and orbitals in the momentum space.

\begin{figure}
\includegraphics[width=87mm]{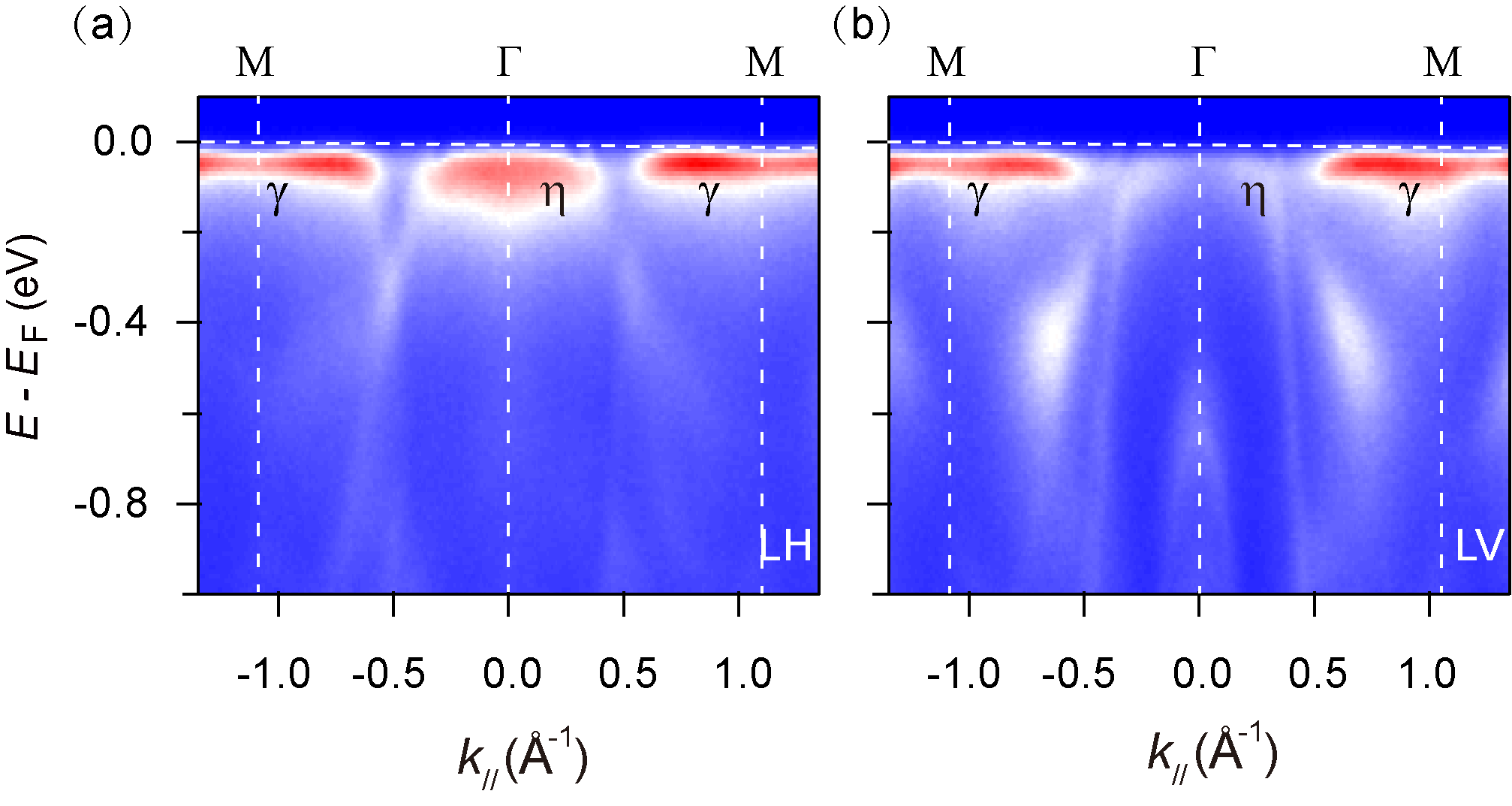}
\caption{(a) Photoemission intensity distributions along $\Gamma$-M taken with LH-polarized light. (b) Photoemission intensity distributions along $\Gamma$-M taken with LV-polarized light.}
\label{band}
\end{figure}
High-quality single crystals of USb$_2$ were grown from Sb flux method with a starting composition of U:Sb = 1:15.  Data of USb$_2$ in Figs. 1 and 2 were obtained at the ``Dreamline" beamline at the Shanghai Synchrotron Radiation Facility (SSRF) with a Scienta DA30 analyzer, and the vacuum was kept below $5\times10^{-11}$~mbar. The overall energy resolution was 16 meV, and the samples were cleaved $in~situ$ along the $c$ axis at 20 K. Data in Figs. 3 and 4 were collected with the in-house ARPES at 10 K, with energy resolution of 12 meV and angular resolution of $0.2^{\circ}$.

Electronic structure calculations for USb$_2$  were performed with a plane-wave basis projected augmented wave method within the density functional theory (DFT) framework, as implemented in the Vienna Ab-initio Software Package (VASP)\cite{Kresse.93}. All 5$f$ electrons of U were treated as Bloch states in the calculations, and the spin-orbit coupling effect was considered using a second variational step. The Perdew-Burke-Ernzerhof (PBE) flavor\cite{Perdew.96} of the generalized gradient approximation was adopted to describe the exchange-correlation of the valence electrons. Based on the optimized bulk antiferromagnetic unit cell, symmetric (001) slabs with Sb- and U-terminations were also constructed. The Sb (II)-terminated slab model consists of 10 atomic layers with 15 ${\AA}$ vacuum layer.  An energy cutoff of 600 eV and $21 \times 21 \times 1$ $k$-mesh in the Monkhorst-Pack scheme \cite{Monkhorst.76} were employed to converge the structure optimization and static self-consistence calculations to better than 1 meV/atom. For the structure relaxation, only outer three atomic layers from each surface of the slab were allowed to move until the residual force on every atom is smaller than 0.01 eV/${\AA}$.


\begin{figure*}[tbp]
\includegraphics[width=180mm]{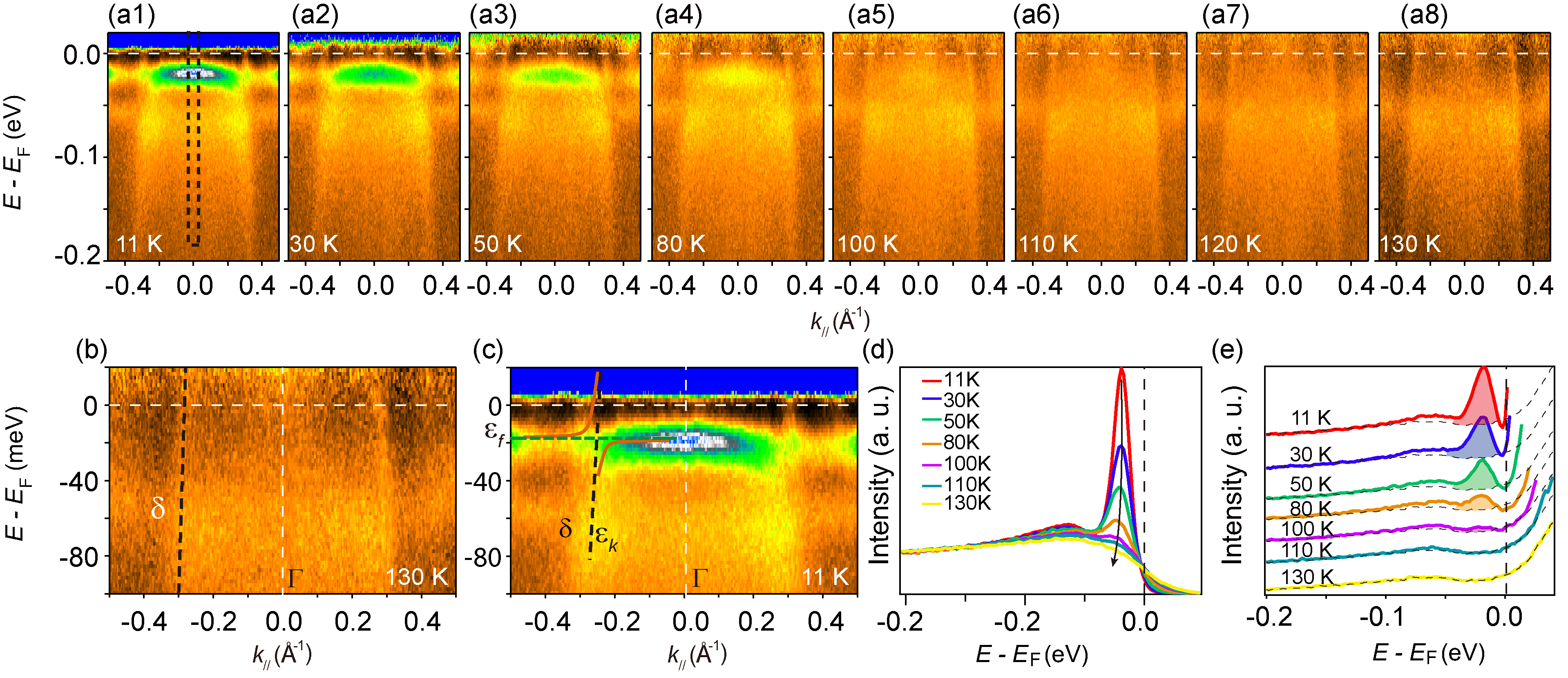}
\caption{Temperature evolution of the electronic structure around the $\Gamma$  point of USb$_2$. (a) Photoemission intensity plot along $\Gamma$-M at the temperatures indicated. The spectra are divided by the resolution-convoluted Fermi-Dirac distribution. (b-c) Zoomed-in ARPES data along $\Gamma$-M at 130 K (b) and 11 K (c).  In panel (c), we present a schematic diagram of the hybridization process under a periodic Anderson model. The black dashed line is the conducting $\delta$ band. The orange curve represents the hybridized band. The green line denotes the position of t  he $f$ band. (d) Temperature dependence of the quasiparticle spectral weight in the vicinity of $\Gamma$. Integrated window has been marked by the black dashed block in (a1). (e) Same as (d) after dividing by the Fermi-Dirac function. The dashed line represents the EDC taken at 130 K, and the shaded regions represent the difference between the low temperature data and 130 K data.}

\label{resonantPES}
\end{figure*}

\begin{figure}
\includegraphics[width=87mm]{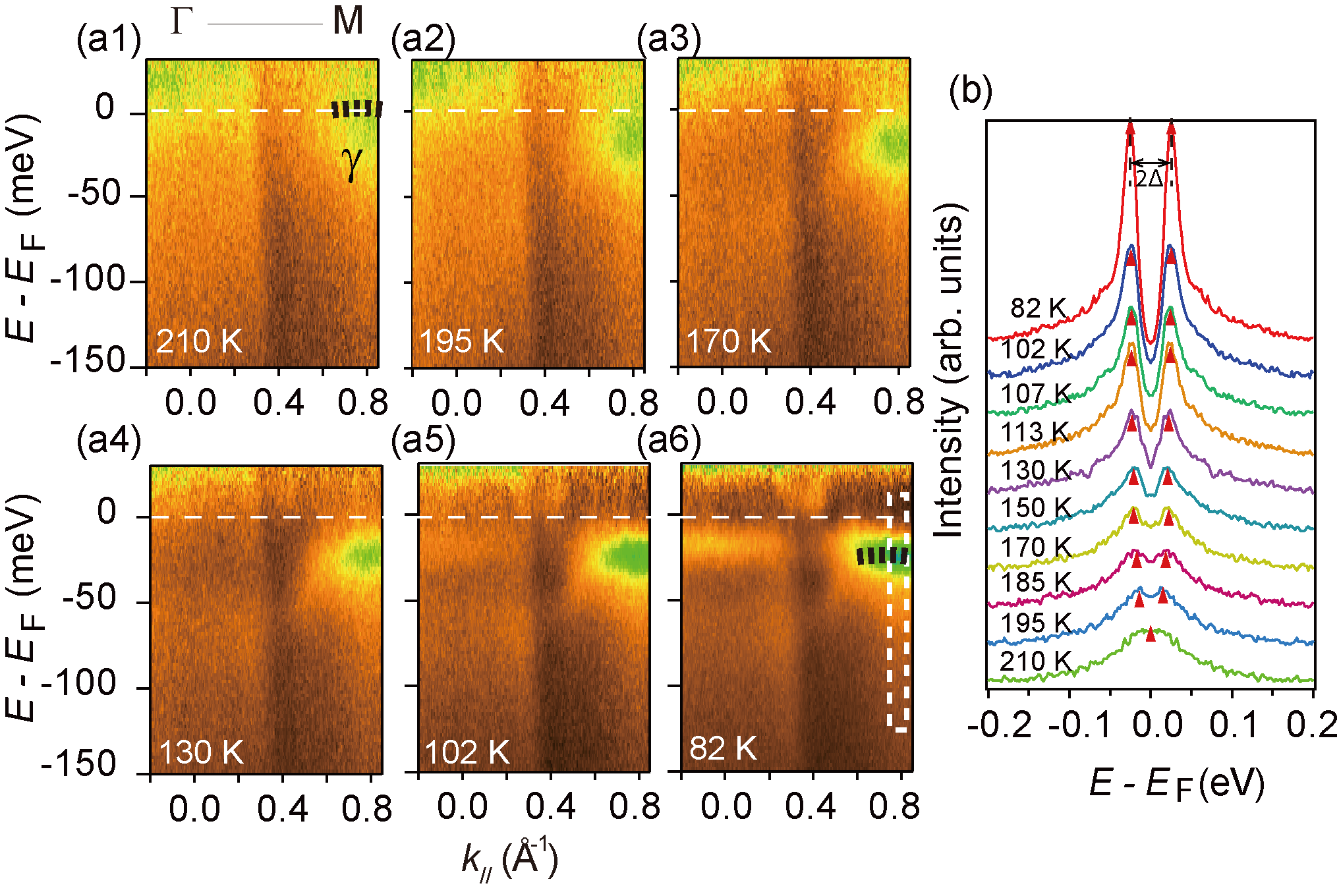}
\caption{Antiferromagnetic transition around the M  point in USb$_2$. (a) Temperature dependence of the band structure along $\Gamma$-M. (b) Temperature dependence of the symmetrized EDCs around the M point, and the EDCs are integrated over the white rectangle area in (a6). }
\label{tdep}
\end{figure}

Detailed characterization of our USb$_2$ samples can be found in Fig. S1 of the supplemental material. Photoemission intensity map of USb$_2$ at 20 K taken with 102 eV LH-polarized photons is displayed in Fig. 1(a). The Fermi surface consists of one diamond Fermi pocket around the zone center, which can be assigned to the $\gamma$ band. Two Fermi pockets $\alpha$ and $\beta$ can be observed around the X point. Interestingly, we find large spectral weight around both the $\Gamma$ and M points, which is due to the formation of heavy quasiparticle bands near $E_F$.
Figs. 1(c-e) show the band structure along several high-symmetry directions. Two nearly flat bands named $\eta$ and $\gamma$ in the vicinity of $E_F$ can be observed around the $\Gamma$ and M point, respectively. Along $\Gamma$-M in Fig. 1(c), only one  band $\gamma$ crosses $E_F$, which contributes the diamond pocket around the zone center. The $\eta$ band hybridizes with the nearly flat $f$ band, forming a heavy quasi-particle band near the $\Gamma$ point. Meanwhile, the nearly flat $\gamma$ band  contributes large spectral weight near $E_F$ around the M point. Detailed electronic structure around the X point can be clearly identified from the band structure along the $\Gamma$-X direction in Fig. 1(e), and two conducting bands $\alpha$ and $\beta$ can be observed crossing $E_F$, forming two Fermi pockets around the X point. The Fermi energy crossings of these bands can also be clearly observed along M-X, see Fig. 1(d).
To identify the cleaved surface for our ARPES measurements, we have performed band structure calculations for both U- and Sb- terminations and found that our ARPES results agree well with the band structure for the Sb-terminated surface in Figs. 1(b) and 1(f). Both the Fermi surface and band structure are in good accordance with experimental results: the observed $\alpha$, $\beta$, $\gamma$ and $\eta$ bands near $E_F$ can all be found in the calculation, and a $\sigma$ band located at much higher binding energy can also be found, which has been marked purple in Fig. 1(f) and can not be reproduced in bulk calculation in literatures \cite{Lebegue.06}.

The nearly flat $\eta$ and $\gamma$ bands around the $\Gamma$ and M points exhibit weak dispersion near $E_F$, which are mainly originated from the U 5$f$ orbitals. More interestingly, we found that the $\eta$ band locates at 17 meV below $E_F$, while the $\gamma$ band around  the M point locates at much higher binding energy of 27 meV. To further identify the orbital character of these bands, we performed photoemission measurements with different polarized photons, which is sensitive to the orbital character of the band structure. Figs. 2(a) and 2(b) display the photoemission intensity distributions along the $\Gamma$-M direction taken with LH- and LV-polarized light, respectively. For the nearly flat $\gamma$ band around  the M point, the intensity of its spectral weight does not change much with different polarization, while for the $\eta$ band around the $\Gamma$ point, $f$ spectral weight near $E_F$ has been greatly reduced with LV-polarized light, comparing with LH-polarization. This dramatically different behavior of the two  bands indicates that they have different orbital character.

In order to reveal the origins of these two nearly flat bands in USb$_2$, we performed detailed temperature-dependent measurements with 21.2 eV  photons both around the $\Gamma$ and M points. Figure 3 displays temperature evolution of the electronic structure around $\Gamma$ after dividing by the Fermi-Dirac function. While at 130 K in Fig. 3(a8), only one conduction band $\delta$ with fast dispersion can be observed. Upon decreasing temperature, weakly dispersive $f$-electron feature gradually emerges with increasing weight near $E_F$. This temperature dependent evolution can be further illustrated by the energy distribution curves (EDCs) around the $\Gamma$ point in Fig. 3(d).  Figs. 3(b) and 3(c) present a detailed comparison of the electronic structure near $E_F$ at 130 and 11 K. At 130 K, only the conducting $\delta$ band can be resolved in Fig. 3(b). At 11 K, an obvious $f$-electron feature near $E_F$ can be clearly observed in Fig. 3(c), which is due to the hybridization between the $f$ band and the conducting $\delta$ band. This results in the redistribution of the $f$ spectral weight and forms the weakly dispersive band near $E_F$.  This hybridization process can be described under the periodic Anderson model, as illustrated in Fig. 3(c).
The $f$-states develop with decreasing temperature, enhancing the $f$-electron spectral weight in Fig. 3(c). Such a process starts from a much higher temperature, and to estimate the onset temperature of the formation of this $f$ band more precisely, we plot the EDCs around the $\Gamma$ point after dividing by the Fermi-Dirac distribution in Fig. 3(e). From our ARPES data, the $f$ band almost disappears at 130 K. Herein, we use the difference between the lower temperature data and 130 K data to determine the onset temperature, and the shaded region in Fig. 3(e) represents this difference. Since the localized-to-itinerant transition of the $f$ electrons is actually a crossover behavior, it is hard to accurately determine the onset temperature. However, difference can be observed at 100 K which gives the onset temperature of 100 $\pm$10 K. The observed onset behavior in USb$_2$ is similar to the evolution of the 4$f$-electron behavior in Ce-based compounds \cite{Qiuyun.Co,Qiuyun.Rh,Qiuyun.Ir,Qiuyun.Ce}. This direct visualization of the hybridization between the $f$ bands and the conduction bands signifies the onset of Kondo correlations between  the local moments with conduction electrons.

USb$_2$ orders antiferromagnetically below a high N$\acute{иж}$eel temperature of 203 K \cite{Wawryk.06,Aoki.99}. To study the evolution of the electronic structure across this phase transition, we performed temperature dependent measurements along $\Gamma$ -M  from 82 up to 210 K and found the $\gamma$ band around the M point changes dramatically.
At 210 K above the transition temperature, some spectral weight can be observed at $E_F$, which was enhanced and shifted away from $E_F$ with lowering temperature, resulting in an energy gap opened at $E_F$. From the symmetrized EDCs of the $\gamma$ band around the M point, we can track the gap opening behavior. This gap opens between 195 and 210 K, and the gap size gradually increases with decreasing temperature and saturates at low temperature. The largest gap size is about 27 meV at 80 K. The opening temperature of the gap is in line with the antiferromagnetic transition temperature (203 K) in USb$_2$, indicating that the formation of the nearly flat $\gamma$ band is related to the magnetic order.

In heavy-fermion compounds, containing 4$f$ or 5$f$ electrons, the interplay of localization and itinerancy is central to understanding their exotic properties. The $f$ electrons are essentially localized in the paramagnetic state at high temperature. As the temperature is lowered, their exchange coupling to conduction electrons leads to the formation of heavy quasi-particles and become itinerant. The competition between the Kondo effect and RKKY interaction determines the ground state, and is a central issue of the field.


In this work, we observed two nearly flat bands in USb$_2$ located at the $\Gamma$  and M points, respectively, which show obviously different behavior with temperature. Polarization-dependent measurements further reveal that they are from different orbitals. Temperature-dependent behavior of the nearly flat band around the $\Gamma$  point is similar to the localized-to-itinerant crossover of 4$f$ electrons in Ce- based compounds \cite{Qiuyun.Co}, which can be assigned to the Kondo interaction of 5$f$ states with conduction electrons in USb$_2$. While for the $\gamma$ band around the M  point, the opening of a gap around 203 K is consistent with the antiferromagnetic transition temperature of USb$_2$. This different temperature-dependent behavior of the two nearly flat bands may in principle arise from two types of orbital selectivity. One is that the conduction electron states near the $\Gamma$ and M points are different. This situation has been recently reported by  Jang $et~al.$  in the Kondo semimetal CeSb \cite{Sooyoung Jang.19}. The other is that the distinction between the $\Gamma$ and M points reflects the involvement of different 5$f$ orbitals. Our results can be explained by the latter of 5$f$-electron orbital selectivity, since the main contribution of these two bands is from the $f$ orbitals,  by comparing with the electronic-structure calculations in Fig. S2 of the supplemental material. This orbital-selective behavior of 5$f$ states manifests itself in different bands in the momentum space, which reconciles the Kondo entanglement with the antiferromagnetic order.



To summarize, our results provide a comprehensive experimental picture of the localized-to-itinerant crossover behavior in actinide compounds, and reveal the electronic structure evolution across the antiferromagnetic transition. More importantly, we find that the orbital selectivity of the 5$f$ electrons facilitates the existence of itinerant 5$f$ electrons in the antiferromagnetic state of USb$_2$.
We expect this new insight to be important for obtaining a complete microscopic understanding of the uranium-based materials and, by extension, the multiple-orbital physics of strongly correlated systems in general.



\begin{acknowledgments}
We gratefully acknowledge helpful discussions with Prof. D. L. Feng, H. Q. Yuan, X. Dai, Y. F. Yang and J. B. Qi. This work is supported by  the National Key Research and Development Program of China (No. 2017YFA0303104), the Science Challenge Project (Grants No. TZ2016004) and the National Science Foundation of China (Grants No. 11874330, U1630248). Work at Rice was in part supported by the NSF Grant No. DMR-1920740 and the Robert A. Welch Foundation Grant No. C-1411.

Q. Y. Chen, X. B. Luo and D. H. Xie contribute equally.
\end{acknowledgments}

\end{document}